\documentclass[preprint,aps,prd,showpacs,nofootinbib]{revtex4}
\usepackage{epsfig}
\usepackage{graphicx}

\newcommand{\br} [1]{ \left( #1 \right) }
\newcommand{\brs}[1]{ \left[ #1 \right] }

\newcommand{\nn}{ \nonumber }

\begin{document}

\title{The structure function method applied to
polarized and unpolarized electron-proton scattering: a solution of the $G_E(p)/G_M(p)$ discrepancy}

\author{Yu. M. Bystritskiy and E. A. Kuraev}
\affiliation{\it JINR-BLTP, 141980 Dubna, Moscow region, Russian
Federation}

\author{E. Tomasi-Gustafsson}

\affiliation{\it DAPNIA/SPhN, CEA/Saclay, 91191 Gif-sur-Yvette
Cedex, France}

\date{\today}

\begin{abstract}
The cross section for polarized and unpolarized electron-proton scattering is
calculated taking into account radiative
corrections in leading and next-to leading logarithmic
approximation. The expression of the cross section is formally similar to
the cross section of the Drell-Yan process, where the structure functions
of the electron play the role of Drell-Yan probability distributions.
The interference of the Born
amplitude with the two photon exchange amplitude (box-type diagrams) is expressed as
a contribution to $K$-factor and it is larger when the momentum is equally shared
between the two photons, assuming that proton form factors decrease rapidly with
the momentum transfer squared. The calculation of the box amplitude is done when
the intermediate state is a
proton or the $\Delta$-resonance. The results of numerical estimations show that
the present calculation of radiative corrections can bring into agreement the
conflicting experimental results on proton electromagnetic form factors an that
the two photon contribution is very small.
\vspace{3cm}

\end{abstract}

\maketitle

\section{Introduction}
Radiative corrections (RC) to elastic (inelastic) electron-proton
($ep$) scattering cross sections can be classified in two types, according
to the reaction mechanism which is assumed:
one where a virtual photon is exchanged between electron and proton and a second
one taking into account the two virtual photon
exchange amplitude, arising from box-type Feynman diagrams in the lowest
order of perturbation theory
(PT). Both kinds of contributions to RC were considered
in the literature, in detail, at the lowest order of PT for polarized and
unpolarized cases.

The most elaborated consideration at the lowest order of PT was done in
Ref. \cite{MaxT}, where the approaches of previous papers (cf. the reference
list in \cite{MaxT}) were considerably improved. The role
of higher orders of PT was firstly considered for the unpolarized case in the
limit of hard photon emission in Ref. \cite{KMF} and later for polarized
case in Refs. \cite{AAM} and \cite{DKSV}.

The elastic $ep$ cross section decreases very rapidly with the momentum transfer
squared, $Q^2=-q^2$, proportionally to $Q^{-4}$. The size of RC essentially depends
on how the experiment was performed. For example, in experiments where only
the angle of the scattered electron is measured, the initial electron emission
can induce an enhancement of RC due to decreasing of $Q^2$.

Initial and final lepton emission can be taken into account by writing the cross section
in form of cross section of Drell-Yan process where the structure
functions of the electron (SFs) play the role of probability distributions
\cite{KMF}. The set of SFs obey the renormalization group equations
(Lipatov's equations). Their solutions are well known
\cite{KF}. The formalism of SFs allows one to obtain the cross
section in the so called "leading logarithmic approximation (LLA)
i.e. taking correctly into account the terms of the order
$[(\alpha/\pi)\ln(Q^2/m_e^2)]^n$. It corresponds to collinear
kinematics, where the photon is emitted in a direction close to the direction
of the electron. Knowing the value of
RC in the lowest order of PT, the non-leading
contribution $(\alpha/\pi)[(\alpha/\pi)\ln(Q^2/m_e^2)]^n$ can be calculated.

A different source of enhancement of cross section is related to the so called
Weizsacker-Williams kinematics, where photons are emitted in non-collinear
kinematics, and provide almost zero momentum $Q^2<m_e^2$. This is not
discussed  in the present work.

A possible enhancement of the elastic cross section can be
associated with the box type Feynman diagram, when the momentum squared is
equally shared between the two photons. The relative contribution of two
photon exchange, from simple counting in
$\alpha$, would be of the order of the fine
structure constant, $\alpha=\displaystyle\frac{e^2}{4\pi}\simeq \frac{1}{137}$:
any contribution of the two-photon exchange (through its interference with
the one-photon mechanism) would not exceed $1\%$. But long ago it was
observed \cite{Gu73} that the simple rule of $\alpha$-counting for the
estimation of the relative role of two-photon contribution to the
amplitude of elastic
electron hadron scattering does not hold at large momentum
transfer. Using a Glauber approach for the
calculation of multiple scattering contributions \cite{Bl71}, it appeared that
the relative role of two-photon exchange can increase significantly in the
region of high momentum transfer, due to the
rapid decreasing of proton form factors(FFs) in the case of
proton intermediate state. The relevant quantity is the square of the FFs,
calculated at $Q^2/4$ and it can at least partially compensate the factor
of $\alpha$. A similar effect takes place when the
$\Delta(33)$ resonance is present in the intermediate state of the box diagram,
because the transition in the vertex $\gamma^*p\Delta$ shows also a
rapid decreasing with $Q^2$ \cite{Fr99}.

Taking a simple model for nucleon form factors, based on the dipole parametrization:
\begin{equation}
G_E(Q^2)=\frac{G_M(Q^2)}{\mu}=G_D(Q^2)=\frac{M_0^4}{(Q^2+M_0^2)^2},~
M_0^2=0.71~{\mbox GeV}^2,~\mu=2.79
\label{eq:eqff}
\end{equation}
when both exchanged photons have momenta close to $q/2$,
an enhancement factor appears in the loop calculation: the ratio of FFs
in Born and Box-type amplitudes. This specific kinematics differs from the "one soft photon"
approach used in the past, Ref. \cite{MaxT}, when considering the box diagram.

Recently, the interest in the $2\gamma$ contribution was revived
as a possible explanation of the discrepancy among experimental data
on elastic $ed$ scattering \cite{Re99}. Electromagnetic proton form factors
show a different behavior as a function of $Q^2$, when measured
with two different methods: the polarization transfer
method \cite{Re68}, which allows a precise measurement
of the ratio of the electric to magnetic proton form
factors \cite{Jo00} and the Rosenbluth separation,
from unpolarized elastic $ep$ cross section \cite{Ar04}.

In Ref. \cite{ET04} it was noted that the reason of the
discrepancy lies in the slope of the reduced cross section
as a function of $\epsilon$, the virtual photon polarization.
At the kinematics of the present experiments, radiative
corrections on the cross section can reach up to $40\%$ ,
and affect very strongly this slope, changing even its
sign, when $Q^2\ge 2$ GeV$^2$.

In Ref. \cite{Bl05} it was claimed that the presence of the two photon
contribution can bring in agreement the data on the proton
electromagnetic form factors from polarization transfer and the
Rosenbluth methods. However, the kinematical properties related
to fast decreasing FFs were not investigated in detail, as well as the
possible presence of inelastic contributions in the intermediate state.
A possible test of the model dependence of the calculation with an exactly
solvable QED result is also absent.

On the other hand, Ref. \cite{AAM} is very detailed.  The SFs
method was applied to transferred polarization experiments. The
size of this effect was an order of magnitude too small to
bring the polarization data in agreement with the unpolarized
ones. Therefore the conclusion of that paper was that one
could not solve the discrepancy among the existing data.

The SF method was also applied to polarization observables
in Ref. \cite{DKSV}, where it was shown that the corrections
can become very large, if one takes into account the initial
state photon emission. However the corresponding
kinematical region is usually rejected in the experimental
analysis, by appropriate selection on the scattered
electron energy.

The motivation of the present paper is the stress the need for present
experiments to go beyond the lowest order of PT, using leading logarithmic
approximation and beyond. RC traditionally applied are proportional to
$\ln(\Delta E)/E\ln(-q^2/m^2)$, where $E$ is the laboratory beam energy,
$q^2$ the momentum transfer squared and $\Delta E$ is the maximum energy
of the undetected photon. In recent experiments $E$ is large and the
experimental resolution is very good  (allowing to reduce $\Delta E$).
Therefore this term becomes sizeable and one can not safely neglect
higher order corrections. A complete calculation of radiative corrections
should take into account consistently all different terms which contribute
at all orders (including the two photon exchange contribution) and their
interference. We derive an expression of the radiative corrected cross
section for $ep$ elastic scattering, in  both polarized and
non-polarized cases, which is easy to handle for
experimentalists and which has a sufficient accuracy.

We will show firstly that the $G_E(p)/G_M(p)$ problem can be solved by
taking into account initial state emission, in SF approach, and, secondly,
that the $2\gamma$ exchange mechanism is irrelevant for the solution of this problem.

Our paper is organized as follows.
In section \ref{DrellYanExpressions} we give the
Drell-Yan formulas for cross sections in polarized and unpolarized
cases. Section \ref{KFactorCalculation}
is devoted to the calculation of the contribution of two photon exchange,
 for the unpolarized cross section and of the degree of
transversal and longitudinal polarization of recoil proton.
Numerical results are presented and discussed in
Section \ref{ResultsAndDiscussion}.
Conclusions summarize the main points of this work. Two appendices
contain details of the calculation.

\section{Drell-Yan expression of the $ep$ cross sections in unpolarized and
polarized cases}
\label{DrellYanExpressions}

It is known \cite{KMF} that the process of emission of hard photons by
initial and scattered electrons plays a crucial role, which results in
the presence of the radiative tail in the distribution on the scattered
electron energy. The structure functions (SF) approach extends the traditional
one \cite{Mo69}, taking precisely into account the contributions of higher
orders of perturbation theory and the role of initial state photon emission.
The cross section is expressed in terms of SF of the initial electron and
of the fragmentation function of the scattered electron  energy fraction.
The dependence of the differential cross section on the
angle and the energy fraction of the scattered electron
$y=1/\rho $ ($\rho$ is the recoil factor $\rho=1+(E/M)(1-\cos\theta)$) can be written as:
\begin{equation}
\frac{d\sigma}{d\Omega dy}=
\int_{z_0}^1\frac{dz\rho_z}{z}D(z)
D\left(\frac{y\rho_z}{z}\right )
\frac{\Phi_0(z)}{|1-\Pi(Q^2_z)|^2}\left (1+\frac{\alpha}{\pi}K \right ).
\label{eq:eqy}
\end{equation}
The term $K=K_e+K_p+K_{box}$ is the sum of three contributions. $K_e$ is
related to non leading contributions arising from the pure electron block
and can be written as \cite{KF,KMF}:
\begin{equation} K_e= -\displaystyle\frac{\pi^2}{6} -\displaystyle\frac{1}{2} -\displaystyle\frac{1}{2}\ln^2\rho+Li_2(\cos^2\theta/2),~
Li_2(z)=-\int_0^z \displaystyle\frac{dx}{x} \ln(1-x).
\label{eq:eqll}
\end{equation}
A second term, $K_p$  concerns proton emission. The emission of virtual and
soft photons by the proton is not associated with large logarithm: $L$,
therefore the whole proton contribution can be included as a $K_p$ factor:
\begin{eqnarray}
K_p&=&\displaystyle\frac{1}{\beta}\left \{ -\displaystyle\frac{1}{2}\ln^2x-\ln x\ln (4(1+\tau))
+\ln x - \right .
\nonumber \\
&&\left . (\ln x-\beta)\ln\left [ \displaystyle\frac{M^2}{4E^2(1-c)^2}\right ]+\beta
-Li_2\left(1-\displaystyle\frac{1}{x^2}\right)+2Li_2\left(-\displaystyle\frac{1}{x}\right)+
\displaystyle\frac{\pi^2}{6} \right\}
\label{eq:eqkp}
\end{eqnarray}
with $x=(\sqrt{1+\tau}+\sqrt{\tau})^2$, $\beta=\sqrt{1-M^2/E'^2}$ and
$E'=E(1-1/\rho)+M$ are the scattered proton velocity and energy.
The contribution of $K_p$ to the $K$ factor is of the order of -2 thousandth for
$c=0.99$, $E=21.5$ GeV, $Q^2$=31.3 GeV$^2$ \cite{MaxT} .

Lastly, $K_{box}$ represents the interference of electron and proton
emission. More precisely the interference between the two virtual photon
exchange amplitude and the Born amplitude and the relevant part of the
soft photon emission i.e., the interference between the electron and proton
soft photon emission, may be both included in the term $K_{box}$. This
effect is not enhanced by large logarithm (characteristic of SF) and can
be considered among the non-leading contributions. It is an $\epsilon$-independent
quantity of the order of unity, which includes all the non-leading terms,
as two photon exchange and soft photon emission.

The non-singlet SF is (the method for using SF in the numerical calculation
is described in Appendix A):
\begin{equation}
D(z,\beta)=\frac{\beta}{2}\brs{\br{1+\frac{3}{8}\beta}(1-z)^{\frac{\beta}{2}-1}-
\frac{1}{2}(1+z)} \br{1+O(\beta)},
\label{eq:eq6}
\end{equation}
\begin{equation}
    \beta=\frac{2\alpha}{\pi}(L-1), \qquad
    Q^2=\frac{2E^2(1-\cos\theta)}{\rho},~~L=\ln\frac{Q^2}{m_e^2},
\label{eq:eq6a}
\end{equation}
$m_e$ is the electron mass. The lower limit of integration, $z_0$, is related
to the 'inelasticity' cut, $c$, necessary to select the elastic data:
\begin{equation}
z_0=\frac{c}{\rho -c(\rho-1)}.
\label{eq:eqz}
\end{equation}

The Born cross section for the scattered electron, $\Phi_0$, is:
\begin{equation}
\Phi_0(Q^2,\epsilon)=\frac{\sigma_M}{\epsilon \rho(1+\tau)}\sigma_{red}(Q^2,\epsilon),
~\sigma_{red}(Q^2,\epsilon)={\tau }G_M^2(Q^2)+\epsilon G_E^2(Q^2),
\label{eq:eqphi}
\end{equation}
where
$\sigma_M=\alpha^2\cos^2(\theta/2)/[4E^2\sin^4(\theta/2)]$
is the Mott's cross section and
\begin{equation}
\tau=\frac{Q^2}{4M^2},~\frac{1}{\epsilon }=1+2(1+\tau)\tan^2(\theta/2).
\label{eq:eqtauz}
\end{equation}
The vacuum  polarization for a  virtual photon with momentum $q$, $q^2=-Q^2<0$,
is included as a factor $1/[1-\Pi(Q^2)]$. The main contribution to this term
arises from the polarization of electron-positron vacuum:
\begin{equation}
\Pi(Q^2)=\frac{\alpha}{3\pi}\left [L-\frac{5}{3} \right ].
\label{eq:eqpi}
\end{equation}

The $z$-dependent kinematically corrected quantities (corrected for the shift
in momentum due to the photon emission) are obtained from the corresponding ones by
replacing the initial electron energy $E$ by $zE$.

\begin{figure}
\begin{center}
\includegraphics[width=8cm]{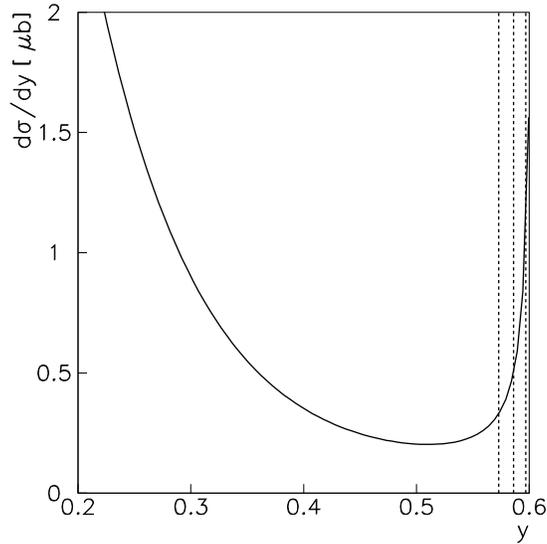}
\caption{The $y$ dependence of the
elastic differential cross section, at
$\theta =32.4^0$ and $Q^2$=3 GeV$^2$.}
\label{Fig:figy}
\end{center}
\end{figure}

The $y$ dependence, at fixed momentum transfer and
electron scattering angle, shows a steep rise, at
small $y$ due to initial state emission, and a rise
in the vicinity of the elastic value, $y=1/\rho$. As
an example, such dependence is shown in Fig. \ref{Fig:figy},
for $\theta =32.4^0$ and $Q^2$=3 GeV$^2$. The dashed
lines show the kinematical cuts corresponding to
$c$=0.95, 0.97 and 0.99, from left to right.

In an experiment, the selection of elastic events
requires a cut in the energy spectrum of the scattered
electron, and one integrates over the events where the
energy of the final electron, $E_1'$, exceeds a threshold value
$E_1'>Ey=Ec/\rho$, $\rho=1+(E/M)(1-\cos\theta
)$, $c<1$
($E$ is the initial electron energy).
Due to the properties of SF method, radiative corrections can be written
 in form of initial and final state emission, although gauge invariance is
conserved. This form obeys the Lee-Nauenberg-Kinoshita theorem, about the
cancellation of mass singularities, when integrating on the the final energy
fraction. This results in omitting the final (fragmentation) SF, i.e., in
replacing the term $(\rho_z/z)D(y\rho_z/z)$ (associated with the final electron emission) by unity.

In Ref. \cite{DKSV} the expressions of the transversal and longitudinal
components of the recoil proton polarization were derived in frame of the
Drell-Yan approach.

The relevant observables,$ \Phi^{SF}_{0,L,T}$, i.e., the unpolarized and
polarized (longitudinally and transversal) cross sections calculated in
frame of the SF method, can be written as:
\begin{equation} \Phi^{SF}_{0,L,T}=\int_{z_0}^1dzD(z)\frac{\Phi_{0,L,T}(z)}{|1-\Pi(Q^2_z)|^2}\left [1+\displaystyle\frac{\alpha}{\pi}K_{0,L,T}\right ],
\label{eq:eqdz}
\end{equation}
where the factor
The factors $K_{0,T,L}$ contain
the contribution of the $2\gamma$ exchange diagrams,
and they are estimated in the dipole approximation for FFs in next section \cite{KF}.

It is convenient to write the polarized and unpolarized cross section, in
frame of the  SF method, in the form of a deviation from the expected Born
expressions (we omit RC of higher order):
\begin{equation}
\Phi^{SF}_{0,L,T}=
\Phi^{Born}_{0,L,T}[1+\Delta_{0,L,T}^{SF}+\frac{\alpha}{\pi}K_{0,L,T}]
\label{eq:csa}
\end{equation}
with
$$\Delta_{0,L,T}^{SF}=
\displaystyle\frac{\alpha}{\pi}\left \{ \displaystyle\frac{2}{3}(L-\displaystyle\frac{5}{3})-
\displaystyle\frac{1}{2}(L-1)
\left [2\ln\left (\displaystyle\frac{1}{1-z_0}\right )-z_0
-\displaystyle\frac{z_0^2}{2}\right ]+\right .
$$
\begin{equation}
\left . \displaystyle\frac{1}{2}\frac{\rho(1+\tau)}{ \Phi^{Born}_{0,L,T}}(L-1)\int_{z_0}^1
\displaystyle\frac{(1+z^2)dz}{1-z}
\left [\displaystyle\frac{ \Phi^{Born}_{0,L,T}(z)}{|1-\Pi(Q_z^2)|^2}-
\displaystyle\frac{ \Phi^{Born}_{0,L,T}(1)}{|1-\Pi(Q^2)|^2}\right ] \right \}.
\label{eq:cs}
\end{equation}
where  $\Phi^{Born}_{0,L,T}(1)=\Phi^{Born}_{0,L,T}$ and $\Phi^{Born}_{0}= \Phi_{0}(Q^2,\epsilon)$ from (\ref{eq:eqphi}),
\begin{equation}
\Phi^{Born}_{T}(Q^2,\epsilon)=-2\lambda\left (\frac{1}{\rho}\right )^2
\displaystyle\frac{\alpha^2}{Q^2}
\sqrt{\frac{\tau}{\tan^2(\theta/2)(1+\tau)}}
G_E(Q^2)G_M(Q^2),
\label{eq:eqT}
\end{equation}
\begin{equation}
\Phi^{Born}_{L}(Q^2,\epsilon)=\lambda\displaystyle\frac{\alpha^2}{2M^2}
\left (\frac{1}{\rho}\right )^2\sqrt{1+\displaystyle\frac{1}{\tan^2(\theta/2)(1+\tau)}}
G_M^2(Q^2),
\label{eq:eqL}
\end{equation}
where $\lambda=\pm 1$ is the chirality of the initial electron;

\section{Calculation of the $K$ factors contribution from the $2\gamma$
exchange}
\label{KFactorCalculation}
\subsection{Proton intermediate state}
\begin{figure}
\begin{center}
\includegraphics[width=12cm]{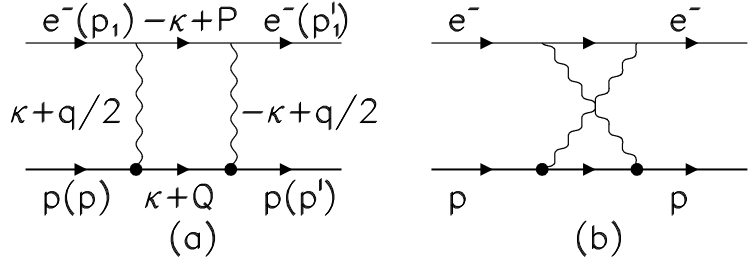}
\caption{\label{Fig:twogam} Feynman diagrams for two-photon
exchange in elastic $ep$ scattering: box diagram (a)
and crossed box diagram (b).}
\end{center}
\end{figure}

The box type Feynman diagram is illustrated in Fig. \ref{Fig:twogam}, where the momenta of the particles are shown in brackets.
Each of the photon carries approximatively half of the transferred momentum $q$. This assumption is justified on the bases of arguments developed in Ref. \cite{Gu73}  and recalled in the Introduction. Let us stress that such approximation leads to an overestimation of the two photon contribution.

\begin{figure}
\begin{center}
\includegraphics[width=10cm]{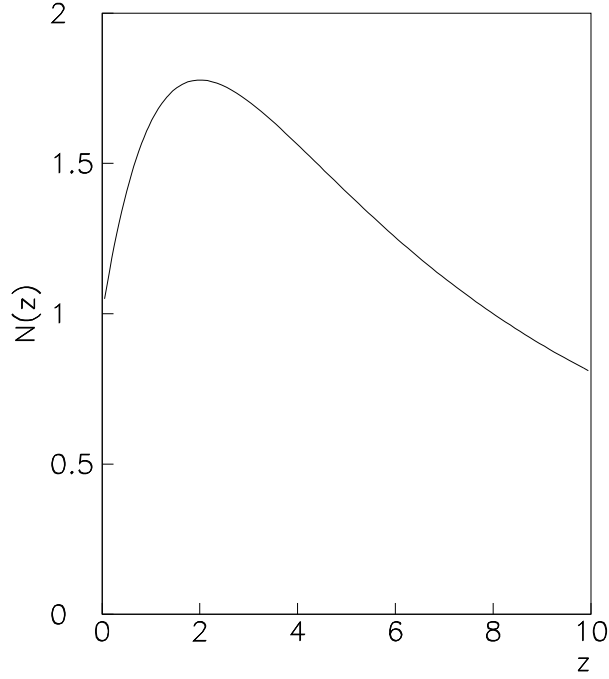}
\caption{\label{Fig:nz} $z$-behavior  of the enhancement factor $N(z)$.}
\end{center}
\end{figure}

We parameterize the loop momentum of the box-type Feynman
amplitude in such a way, that the denominators of Green function are
 $(\pm\kappa+q/2)^2$ for the photon, whereas for the electron $(e)$
and the for the proton $(p)$ they have a form
$(e)=(\pm\kappa+{\cal P})^2-m_e^2$, $(p)=(\kappa+{\cal Q})^2-M^2,$ with ${\cal P}=\displaystyle\frac{1}{2}(p_1+p_1')$, ${\cal Q}=\displaystyle\frac{1}{2}(p+p')$.
The sign '$-$' for the electron corresponds to the Feynman diagram
for the two photon box (Fig.\ref{Fig:twogam}a)  and the sign
'$+$' corresponds to the crossed box diagram (Fig. \ref{Fig:twogam}b).
The assumption of a rapid decreasing of form factors implies
that we can neglect the dependence on the loop momentum
$\kappa$ in the denominators of the photon Green function as well as in the
arguments of the form factors. This results in ultraviolet
divergences of the loop momentum integrals. Therefore
they should be understood as convergent integrals with the cut-off
restriction $|\kappa^2|~<~M^2\tau$:
\begin{eqnarray}
\int\displaystyle\frac{d^4\kappa}{i\pi^2}
\frac{N_\pm({\cal P}, {\cal Q})}
{\br{\br{\pm \kappa+{\cal P}}^2 - m_e^2}
 \br{\br{\kappa+{\cal Q}}^2 - M^2}}
\theta\br{M^2\tau - |\kappa^2|} = I_\pm \cdot N_\pm({\cal P}, {\cal Q}),
\end{eqnarray}
where ${\cal P} = \frac{1}{2}\br{p_1+p_1'}$, ${\cal Q} = \frac{1}{2}\br{p+p'}$.
The explicit form of $I_\pm$ is given in Appendix B.
$N_\pm({\cal P}, {\cal Q})$ is the Feynman diagram numerator defined in Eq. (\ref{UUnp}).
The virtual photons Green function is written as:
\begin{eqnarray}
\displaystyle\frac{1}{\left |\displaystyle\frac{q}{2}\pm \kappa\right |^2}<
\frac{1}{{\cal Q}^2}= \frac{1}{M^2\left(1+\tau\right)}.
\end{eqnarray}
Then the expressions for $K$-factors can be written as:
\begin{eqnarray}
K_i &=& -2 {\cal N}(z)\frac{U^i({\cal P}, {\cal Q})}{\Phi_i}, \qquad i=0,~ L,~ T,
    \label{KFactor}\\
{\cal N}(z)&=&(z+1)^2/[(z/4)+1]^4, \qquad z=Q^2/M_0^2,   \label{nenh}
\end{eqnarray}
where ${\cal N}(z)$ is due to the  dipole dependence of the FFs which is extracted as an
enhancement factor (see Fig. \ref{Fig:nz}). One can see that this factor has a
maximum  equal to $\simeq$2 for $z\simeq 2$, which corresponds to $Q^2\simeq$ 1.4 GeV$^2$.
This behavior is consistent with the results of a rigorous QED calculation \cite{bvv}.
This calculation, which applies to $e\mu$ scattering, gives an upper limit of the 2$\gamma$
contribution to $ep$ scattering, when the muon mass is replaced by the proton mass.

The Born terms, ${\Phi}_i$, $i=0,$ $L$, $T$ have been singled out in the definition of the $K$-factor (see Eqs. (\ref{eq:cs}), (\ref{eq:eqL}), (\ref{eq:eqT})).

In the unpolarized case the expression for $U_0({\cal P}, {\cal Q})$ is:
\begin{eqnarray}
    U_0({\cal P}, {\cal Q}) &=&
    \displaystyle\frac{\alpha^2G_D^2(Q^2)}{16 M^8\rho^2\tau(\tau+1)^2}
    \displaystyle\frac{1}{4} Tr \left [ (\hat{p}'+M)
    \Gamma_\lambda\left (\displaystyle\frac{q}{2}\right ) (\hat {\cal Q} +M)
    \Gamma_\eta \left (\displaystyle\frac{q}{2} \right )
     \br{\hat{p}+M}\bar{\Gamma}_\mu (q)\right ]
    \times \nn \\
    &&\times
    \left\{
           I_+ \cdot
           \frac{1}{4} Tr\brs{\hat{p}_1'\gamma_\lambda \hat{{\cal P}} \gamma_\eta\hat{p}_1\gamma_\mu}
           +
           I_- \cdot
           \frac{1}{4} Tr\brs{\hat{p}_1'\gamma_\eta \hat{{\cal P}} \gamma_\lambda \hat{p}_1\gamma_\mu}
    \right\}\nn \\
    && = N_+I_+ + N_-I_-
    \label{UUnp}
\end{eqnarray}
where
$$\Gamma_\alpha (q)=\frac{F_1(Q^2)}{G_D(Q^2)} \gamma_\alpha -\frac{1}{4M}\frac{F_2(Q^2)}{G_D(Q^2)}[\gamma_\alpha,\hat q], ~
\bar{\Gamma}_\alpha (q)=\frac{F_1(Q^2)}{G_D(Q^2)} \gamma_\alpha +\frac{1}{4M}\frac{F_2(Q^2)}{G_D(Q^2)}[\gamma_\alpha,\hat q], $$
where $F_1(Q^2)$ and $F_2(Q^2)$ are the Pauli and Dirac FFs, related to the Sachs FFs by
$$F_1(Q^2)=\frac{G_E(Q^2)+\tau G_M(Q^2)}{1+\tau},
~F_2=\frac{G_M(Q^2)-G_E(Q^2)}{1+\tau}.$$

The quantities $U_{T,L}({\cal P}, {\cal Q})$ for polarized case can be obtained from
(\ref{UUnp}) by the following replacements:
\begin{eqnarray}
    \gamma_\mu \to \gamma_\mu\gamma_5 \label{LeptonSpurReplacements}
\end{eqnarray}
in the lepton traces and
\begin{eqnarray}
    \br{\hat{p}'+M} \to \br{\hat{p}'+M} \hat{a}_{T,L} \gamma_5
     \label{ProtonSpurReplacements}
\end{eqnarray}
in the proton traces. Here $a_{T,L}$ is the final proton polarization vector
(i.e. $(a_{T,L} p') = 0$) and corresponds to different
orientations of the proton polarization. If the final
proton is polarized along the $x$-axis, one finds:
\begin{eqnarray}
\br{a_T p} = 0, \qquad
\br{a_T p_1} = -\frac{E^2}{2M\rho}\frac{\sin\theta}{\sqrt{\tau(1+\tau)}},
\end{eqnarray}
whereas in case of polarization along the $z$-axis:
\begin{eqnarray}
\br{a_L p} = 2 M \sqrt{\tau(1+\tau)}, \qquad
\br{a_L p_1} = M \sqrt{\frac{\tau}{1+\tau}}\br{\frac{E}{M}-1-2\tau}.
\end{eqnarray}

\subsection{The $\Delta$ resonance contribution}
Let us write the structure of the vertex for the transition
$\Delta(p)\to \gamma^*(q)+P(p')$, following the formalism
of Ref. \cite{Ko05,AR77} (and references therein):
\begin{eqnarray}
M(\Delta \to \gamma^* P) &=& e g_{\Delta N}~
\sqrt{3/2}~\bar{u}(p', \eta)\br{\gamma_\mu - \frac{1}{M_\Delta}p'_\mu}u_\nu(p, \lambda)
\displaystyle\frac{F_{\mu\nu}(q)}{ 2\sqrt{Q^2}}G_D(Q^2),
\label{eq:eq18}
\end{eqnarray}
where $F_{\mu\nu}(q)=e_\mu(q) q_\nu-e_\nu(q) q_\mu$ is the Maxwell tensor,
$e(q)$ is the polarization vector of virtual
photon, $\eta$ and $\lambda$ are the chiral states of the nucleon and of the
$\Delta$-resonance and $\sqrt{3/2}g_{\Delta p}\approx 1.56\mu$
($\mu$ is the anomalous magnetic moment of the proton). Moreover, the factor $G_D(Q^2)$ is explicitly extracted.

The Green function of the $\Delta$ resonance, neglecting its width, is
\begin{eqnarray}
-i\frac{D_{\mu\nu}(p)}{p^2-M^2_\Delta+i0},
\label{eq:eq19}
\end{eqnarray}
with
\begin{eqnarray}
    D_{\mu\nu}(p) &=& \sum_\lambda u_\mu(p, \lambda)\bar{u}_\nu(p, \lambda) = \nn\\
    &=& \left (\hat{p}+M_{\Delta}\right )
    \left [g_{\mu\nu}- \frac{1}{3}\gamma_\mu\gamma_\nu-
            \frac{1}{3p^2} (\hat p\gamma_{\mu}p_{\nu}+p_{\mu}\gamma_{\nu}\hat p)\right ],
\end{eqnarray}
$$~D_{\mu\nu}(p)p_{\mu}= D_{\mu\nu}(p)p_{\nu}=0.$$

The transition vertexes associated with form factors
are of the same form as the dipole ones for the nucleons. The part of
the virtual Compton scattering of the proton amplitude which enters in the box
amplitude is:
\begin{eqnarray}
\bar{u}(p')[p']_\mu D_{\rho\sigma}(p_2)[p]_\nu
u(p)F_{\mu\rho}(k_1)F_{\sigma\nu}^*(k_2), \nn
\end{eqnarray}
\begin{eqnarray}
k_{1,2} = \pm\kappa + \frac{q}{2}, \quad
p_2=\kappa + {\cal Q}, \quad
[p]_\mu=\gamma_\mu - \frac{1}{M_{\Delta}}p_\mu, \nn
\end{eqnarray}
where $F_{\mu\nu}(k)=k_\mu e_\nu(k)-k_\nu e_\mu(k)$ is the Maxwell tensor and $e_\mu(k)$ is the photon polarization vector.
Thus, in unpolarized case, the contribution of the
$\Delta$-resonance to the  K-factor :
\begin{equation} K^\Delta_{0}= -2{\cal N}(z)
\displaystyle\frac{U^\Delta_{0}}{\Phi_{0}},
\label{eq:eqdk}
\end{equation}
with

\begin{eqnarray}
    U^\Delta_{0} &=&
   - \displaystyle\frac{\alpha^2G_D^2(Q^2)(1.56~\mu)^2}{64 M^{10}\rho^2\tau^2(1+\tau)^2 }
    \frac{1}{4} Tr\brs{\br{\hat{p}'+M} [p']_\mu D_{\rho\sigma}({\cal Q}) [p]_\nu \br{\hat{p}+M}\bar{\Gamma}_\eta(q)}
    \times \nn \\
    &\times&
    \left\{
        I_+ \cdot \frac{1}{4} Tr\brs{\hat{p}_1' P^{\mu\nu\rho\sigma} \hat{p}_1\gamma^\eta}
        +
        I_- \cdot \frac{1}{4} Tr\brs{\hat{p}_1' R^{\mu\nu\rho\sigma} \hat{p}_1\gamma^\eta}
    \right\},
    \label{DeltaContribution}
\end{eqnarray}
where
\begin{eqnarray}
    P_{\mu\nu\rho\sigma} &=& \frac{1}{4}\brs{\gamma_\rho q_\nu - \gamma_\nu q_\rho}
                             \hat{\cal P} \brs{\gamma_\sigma q_\mu - \gamma_\mu q_\sigma}, \nn \\
    R_{\mu\nu\rho\sigma} &=& \frac{1}{4}\brs{\gamma_\sigma q_\mu - \gamma_\mu q_\sigma}
                             \hat{\cal P} \brs{\gamma_\rho q_\nu - \gamma_\nu q_\rho}. \nn
\end{eqnarray}

The contributions in the polarized cases can be obtained
from (\ref{DeltaContribution})
via the same replacement rules (\ref{LeptonSpurReplacements}),
(\ref{ProtonSpurReplacements}), and the corresponding denominators, $\Phi_{L,T}$.

\section{Results and Discussion}
\label{ResultsAndDiscussion}

\begin{figure}
\begin{center}
\includegraphics[width=10cm]{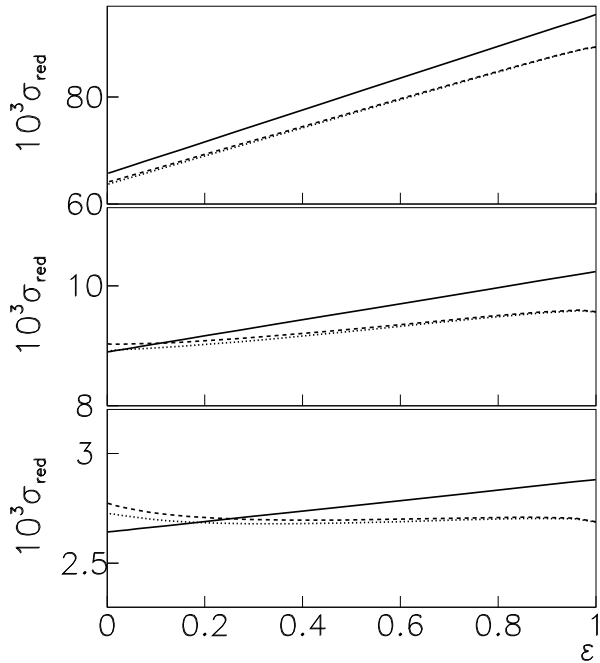}
\caption{\label{Fig:fig3} The $\epsilon$-dependence of the
elastic differential cross section, for $Q^2=1$, 3, and
5 GeV$^2$, from top to bottom: Born cross section (solid line),
Drell-Yan cross section (dashed line), full calculation (dash-dotted line) (including Born, SF and K-factor contributions).}
\end{center}
\end{figure}

The numerical results strongly depend on the experimental conditions, in
particular on the inelasticity cut of the
scattered electron energy spectrum. The results shown here correspond
to $c=0.97$. This value has been chosen because it corresponds to the
energy resolution of modern experiments. The unpolarized cross
section has been calculated assuming the dependence of form
factors on $Q^2$ given by Eq. (\ref{eq:eqff}). In Fig. \ref{Fig:fig3}
the results are shown as a function of $\epsilon$,
for $Q^2=1$, 3, and 5 GeV$^2$, from top to bottom. The
calculation based on the structure function method is shown as dashed lines.
The full calculation,
including the two-photon exchange contribution is shown as
dash-dotted lines. For comparison the results corresponding
to the Born reduced cross section are shown as solid lines.
One can see that the main effect of the present calculation
is to modify and lower the slope of the reduced cross section.
This effect gets larger with $Q^2$. Non-linearity effects are
small and induced by the $y$ integration. Including two-photon exchange modifies very little
the results, in the kinematical range presented here.

\begin{figure}
\begin{center}
\includegraphics[width=10cm]{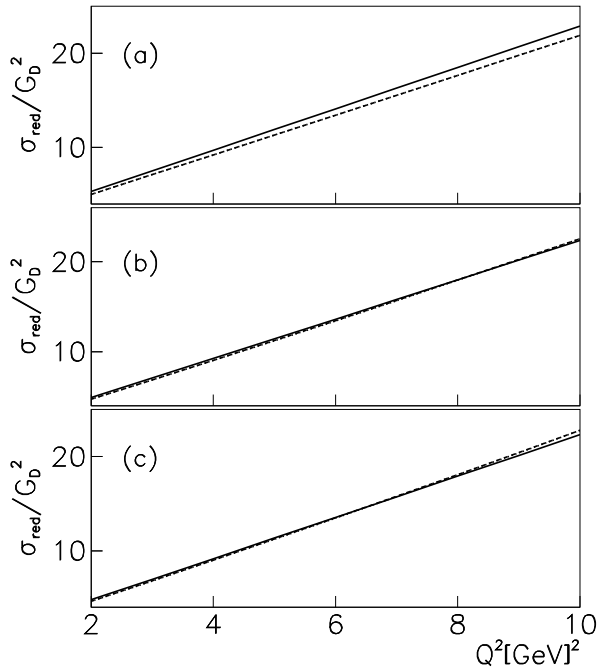}
\caption{\label{Fig:fig4} The $Q^2$-dependence of the
elastic differential cross section, at $\theta =85^0$,
$60^\circ$, and $20^\circ$. Notations as in Fig. \protect\ref{Fig:fig3}.}
\end{center}
\end{figure}

The $Q^2$-dependence of the unpolarized cross section is
shown in Fig. \ref{Fig:fig4}, for  electron scattering
angles equal to $\theta=85^\circ$, $60^\circ$, and $20^\circ$ ,
from top to bottom. The $G_D^2(Q^2)$-dependence has been
removed, in order to enhance visually the differences among the calculations. In spite of this,  the corrections affect very little the  $Q^2$ dependence of the reduced cross section.

\begin{figure}
\begin{center}
\includegraphics[width=10cm]{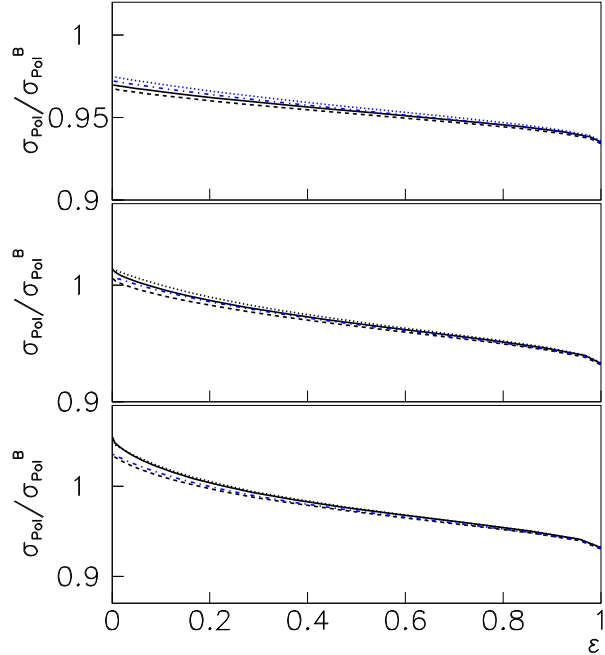}
\caption{\label{Fig:fig5} The $\epsilon$-dependence of the ratio of the  polarized cross section, corrected by the SF method, to the corresponding component of the Born cross section, at  $Q^2=1$, 3, and
5 GeV$^2$, from top to bottom. The calculation including (not including) K-factor is shown as solid line (dotted) for the transversal component  and dashed line (dash-dotted) for the longitudinal component. }
\end{center}
\end{figure}

The results for the polarized case are shown in Figs.
\ref{Fig:fig5} for the
longitudinal (dashed lines) and the transversal (solid lines) components of the
cross section. The ratio of the polarized cross section, corrected with the SF method (Eqs. (\ref{eq:csa},\ref{eq:cs}) to the Born polarized cross section  (Eqs. (\ref{eq:eqL},\ref{eq:eqT}) is reported as a function of $\epsilon$.

The relative effect of the corrections is of the order of the corrections to the unpolarized cross section and is very similar for the two polarized components. This means that the ratio of the polarizations is very little affected by radiative corrections.

Again the effect of the two photon contribution is
negligible, in both cases.
In the specific kinematics considered here, the box type contribution $K_{box}$ does not depend on the soft photon emission parameter, $\Delta E/E$. This quantity is of the order of the $\Delta E$ independent contribution to the charge asymmetry $\Xi( E/M,\cos\theta)$ (corrected by the factor $N(Q^2/M_0^2)$) calculated for $e\mu$ elastic scattering in frame of pure QED (see Fig. 2  of Ref. \cite{bvv})). In the dipole approximation, the form factors reduce to $F_1(Q^2)\to G_D(Q^2)$ and $ F_2(Q^2)\to 0$ when $Q^2$ is large.

\begin{figure}
\begin{center}
\includegraphics[width=10cm]{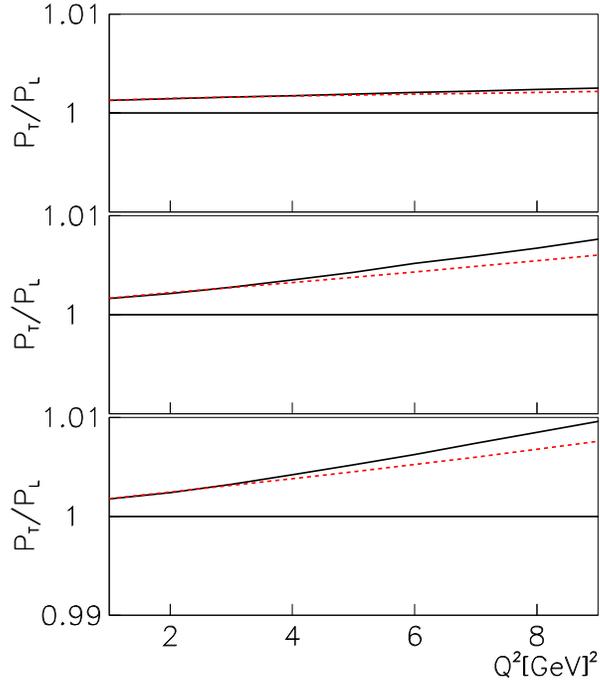}
\caption{\label{Fig:fig6} $Q^2$-dependence of the ratio of the
transversal to longitudinal components of the proton
polarization calculated in the SF method (dashed line), normalized to the same ratio in the Born approximation at $\theta =85^0$, $60^\circ$, and $20^\circ$.}
\end{center}
\end{figure}

It is particularly interesting to look at the ratio of
the transverse to the longitudinal components of the proton
polarization,  which is directly related to the form
factor ratio,  experimentally measured (Fig. \ref{Fig:fig6}, dashed line).
The results are presented after normalization to the Born ratio, in order to compensate the kinematical factors. The correction to be applied to the experimental data, as predicted by the present calculation, is very small, within  1\% at different $\theta$ values, for $Q^2$ up to 10 GeV$^2$. However
the two photon contribution depends on $Q^2$ and becomes larger as $Q^2$ increases.

\begin{figure}
\begin{center}
\includegraphics[width=10cm]{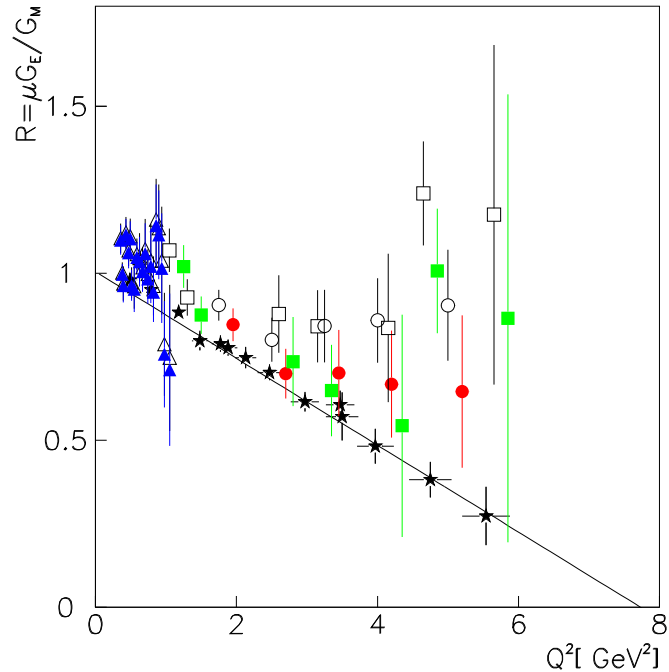}
\caption{\label{Fig:fig8}The $Q^2$-dependence of the FF ratio. Data from Rosenbluth method before (open symbols) and after (solid symbols) correction (c = 0.97), from Ref. \protect\cite{Ja66} (triangles), from Ref. \protect\cite{An94} (squares), from Ref. \protect\cite{Ch04} (circles)  . Data from polarization method \protect\cite{Jo00} are also shown (stars). The line is a fit to the polarization data.}
\end{center}
\end{figure}

One can tentatively extract the FFs ratio, after correcting the measured unpolarized cross section by the ratio of the Born to the SF results. It is evidently an average correction, which is not applied event by event, but it takes into account the main feature of the SF calculation, i.e., the lowering of the slope of the unpolarized cross section as a function of $\epsilon$.
We neglect the correction from the two photon exchange, as it is negligible when compared to the high order corrections.
The $Q^2$-dependence of the ratio $R=\mu G_E/G_M$ is plotted in Fig. \ref{Fig:fig8}, for the sets of data for which a detailed information on RC was published \cite{An94,Ch04} (squares and circles, respectively). Open symbols refer to the published data, the corresponding solid symbols represent the corrected data.
For comparison, a set of data at low $Q^2$,  is also shown (triangles) \cite{Ja66}. Here RC are small, and high order corrections affect very little the results. Data from the polarization method  are shown as stars. Although the corrections would have the effect of getting a larger ratio, the difference between points before and after correction is at most 1\% and can not be seen on this plot. The line corresponds to a fit to polarization data  (for $Q^2\ge$ 1 GeV$^2$) \cite{Jo00}:
\begin{equation}
R(Q^2)=1-(0.130\pm 0.005)\{Q^2~[\mbox{GeV}^2]-(0.04\pm 0.09)\}.
\label{eq:brash}
\end{equation}
The present results suggest that an appropriate treatment of
radiative corrections constitutes the solution of the
discrepancy between form factors extracted by the Rosenbluth
or by the recoil polarization method.

\section{Conclusion}

We have considered radiative corrections in case of
quasi elastic kinematics, when the scattered electron
has energy close to the elastic value. We considered
two types of corrections: the real photon emission related
to the electron vertex, that we calculated  in frame of
the structure function approach, in leading and next to leading orders, the last ones expressed in terms of $K$-factor. We do not include the contribution to $K$-factor due to proton emission, as it is known to be small. The enhancement
of RC has been explicitly calculated in QED, due to emission from the initial lepton.

The contribution to $K$-factor from the interference of electron and proton emission ($2\gamma$- exchange) was estimated in the approximation of fast decreasing of nucleon FFs and found to be small, not exceeding 1\% for the unpolarized cross section. Its contribution is different for the polarized
cross section, and very small on the ratio of the longitudinal
to transversal components.

The K-factor induces a large deviation for small $\epsilon$ values. Such behavior is not physical, but it is due to the approximation used for calculating the box diagram (see Appendix B). The present work is focused on the effect of higher order corrections to the cross section, which are responsible for the largest deviation from the $\epsilon$ dependence of the Born cross section. The non leading contributions are calculated under the assumption that the momentum transfer is almost equally shared between the two exchanged photon. This assumption holds in the limit of a specific kinematics which emphasizes the role of the two photon contribution, at large $Q^2$.

The present results are consistent with a rigorous QED calculation of the $2\gamma$ contribution from Ref. \cite{bvv}, where the process $e^+e^-\to \mu^++\mu^-$ and
the crossed process were calculated and shown to be of the order of 1\%. The QED case can be considered as an upper limit for $2\gamma$ effects in electron proton scattering, due to two reasons, at least: the proton electromagnetic FFs $F_1$ and $F_2$ are smaller than unity. Moreover the contributions of the elastic and inelastic channels, in the intermediate state, should compensate. This has already been pointed out in the literature \cite{Ko05}. We underline that the $\Delta$ resonance contribution should be considered as a model for all possible intermediate states, including $\pi N$, $\pi NN$... Using the analytical properties of the Compton scattering amplitude
\cite{Ku06}, one can espect a large cancellation of the elastic intermediate state (nucleon) and inelastic ones up to a level of 10\% . A complete evaluation of the box diagram for $ep$ elastic scattering, in framework of an analytical model is in preparation \cite{Bytev}.

The main effect of the present calculation of RC is visible on
the unpolarized cross section: it changes noticeably the slope
of the $\epsilon$ dependence of the reduced cross section, in
comparison with the Born approximation. This slope is directly
related to the electric form factor, therefore applying RC as
suggested here to the unpolarized cross section, would solve
the discrepancy between form factors extracted from the Rosenbluth
method and from the recoil polarization method.

In \cite{DKSV} it was shown that the corrections on the
polarization observables can be very large, if the cut parameter
is small, see Fig. \ref{Fig:figy}. This is due to the initial
state photon emission, which is normally excluded in the
experimental analysis.
In this paper we considered the region near the elastic peak
where the contribution to polarized cross section ratio becomes small
(of order 1 \%, see Fig. \ref{Fig:fig6}).

An average SF correction was applied to the data, and significantly improves the consistency of the different sets. However, as pointed out in \cite{ETG04},
this procedure, is still applied as a global factor depending on the relevant variables, $\epsilon$ and $Q^2$ and does not solve the problem of the strong correlation between the parameters of the Rosenbluth fit. An event by event analysis, based on SF method, should be done at the data processing level. This is outside the purpose of this paper.

In conclusion, the SF method can be successfully applied to
calculate RC to elastic $ep$ scattering. In particular,
it takes precisely into account collinear photon emission.
The two photon contribution is negligible in the considered
kinematical range. The correction to the ratio of longitudinal
to transverse proton polarization is small. But the
correction on the unpolarized cross section has the effect
and the size required to solve the discrepancy among proton
form factors.

\begin{acknowledgments}
The authors are thankful to S. Dubnicka for careful reading of the
manuscript and valuable discussions. One of us (E. A. K.) is grateful
to the Institute of Physics, Slovak Academy of Sciences,
Bratislava for warm hospitality and to DAPNIA/SPhN, Saclay,
where part of this work was done.
\end{acknowledgments}

\appendix

\section{Method for the integration of the $D$ function}
\label{AppendixDfunction}

Let us consider the integral
\begin{equation}
{\cal I}=\int_{x_0}^1D(x)\phi(x) dx.
\label{eq:eqa1}
\end{equation}
The partition function   $D(x)$ has a $\delta(x-1)$-type behavior for $x=1$
and has the following properties:
\begin{equation}
\int_0^1D(x)dx =1;~ D(x) |_{x\ne 1}=\displaystyle\frac{\beta}{4}\displaystyle\frac{1+x^2}{1-x} + {\cal O}(\beta^2).
\label{eq:eqa2}
\end{equation}
Therefore Eq. (\ref{eq:eqa2}) becomes:
\begin{eqnarray}
{\cal I}&=& \int_{x_0}^{1-\epsilon} dx D(x)\phi(x)+\int_{1-\epsilon}^1 dxD(x)\phi(1).
\nonumber\\
&=& \displaystyle\frac{\beta}{4}\int_{x_0}^{1-\epsilon}dx\displaystyle\frac{1+x^2}{1-x}\phi(x) +\left (1-\int_{0}^{1-\epsilon}dx\displaystyle\frac{\beta}{4}\displaystyle\frac{1+x^2}{1-x}
\right )\phi(1)+{\cal O}(\beta^2).
\label{eq:eqa2a}
\end{eqnarray}

After elementary integration, Eq. (\ref{eq:eqa2a}) becomes:
\begin{eqnarray}
{\cal I}&=& \displaystyle\frac{\beta}{4}\int_{x_0}^{1-\epsilon}dx\displaystyle\frac{1+x^2}{1-x}
\left [\phi(x)-\phi(1)+\phi(1)\right ]+ \phi(1)
\left [1-\displaystyle\frac{\beta}{4}\int_0^{1-\epsilon}dx\displaystyle\frac{1+x^2}{1-x}
\right ]
\label{eq:eqa5}\\
&=&\phi(1)\left [ 1-\displaystyle\frac{\beta}{4}\left (2\ln \displaystyle\frac{1}{1-x_0}-x_0-\displaystyle\frac{x_0^2}{2}\right )\right ]
+\displaystyle\frac{\beta}{4}\int_{x_0}^1 dx\displaystyle\frac{1+x^2}{1-x}\left [\phi(x)-\phi(1)\right ]+{\cal O}(\beta^2),
\nonumber\\
\end{eqnarray}
removing therefore the singularity.

\section{Calculation of $I_\pm$}
\label{AppendixCalculationIpm}
In this appendix we perform the following integration:
\begin{eqnarray}
    I_\pm = Re \int\frac{d^4\kappa}{i\pi^2}
    \frac{\theta\br{M^2\tau - |\kappa^2|}}
    {\br{{\cal P}_\pm} \br{{\cal Q}}},
     \label{IntegralI1}
\end{eqnarray}
\begin{eqnarray}
    \br{{\cal P}_\pm} &=& \br{\pm \kappa+{\cal P}}^2 - m_e^2, \nn \\
    \br{{\cal Q}} &=& \br{\kappa+Q}^2 - M^2, \nn
\end{eqnarray}
where ${\cal P} = \frac{1}{2}\br{p_1+p_1'}$, ${\cal Q} = \frac{1}{2}\br{p+p'}$.
First we perform Wick-rotation ($\kappa_0 \to i \kappa_0$) and
imply the cut-off provided by $\theta$-function using that parameterizing
\begin{eqnarray}
    Re \int\frac{d^4\kappa}{i\pi^2} \frac{\theta\br{M^2\tau - |\kappa^2|}}
    {\br{{\cal P}_\pm} \br{{\cal Q}}}
    =
    \frac{2}{\pi}
    \int\limits_{-M\sqrt{\tau}}^{M\sqrt{\tau}} d \kappa_0
    \int\limits_0^{M\sqrt{\tau-k_0^2/M^2}} d k~k^2
    \int\limits_{-1}^1 d(\cos \theta_\kappa)
    Re \frac{1}{\br{{\cal P}_\pm} \br{{\cal Q}}},
\end{eqnarray}
where $ k = \left| \vec \kappa \right|$. We also performed the integration over
the azimuthal-angle $\phi_\kappa$.
Now let us consider the integral in the Breit-system where $q_0 = 0$ and
$\vec p_1 = -\vec p_1'$. Thus $\vec {\cal P} = 0$,
$p_0 = p'_0 = E'$, $|\vec p_1| = M\sqrt{\tau}$,
$\vec {\cal Q}^2 = M^2 \cot^2 \br{\theta_e/2}$,
$E' = M \sqrt{\tau + 1/\sin^2\br{\theta_e/2}}$,
where $\theta_e$ is the electron scattering angle in laboratory frame.

Before integrating over angle $\theta_\kappa$ let us write the
explicit expression for real part of integrand:
\begin{eqnarray}
    Re \frac{1}{\br{{\cal P}_\pm} \br{{\cal Q}}}
    =
    \frac{a \br{a+b \cos \theta_\kappa} \mp \delta_1\delta_2}
    {\br{a^2 + \delta_1^2}
    \br{\br{a+b \cos \theta_\kappa}^2+\delta_2^2}}, \nn
\end{eqnarray}
where
$a = -\kappa_0^2 - k^2 + M^2 \tau$,
$b = -2 k |\vec {\cal Q}|$,
$\delta_1 = 2 \kappa_0 M \sqrt{\tau}$.
$\delta_2 = 2 \kappa_0 E'$.
The integration over $\theta_\kappa$ is straightforward and results in:
\begin{eqnarray}
    I_\pm &=& -\frac{1}{\pi |\vec {\cal Q}|}
    \int\limits_{-M\sqrt{\tau}}^{M\sqrt{\tau}} d \kappa_0
    \int\limits_0^{M\sqrt{\tau-k_0^2/M^2}} d k~k
    \frac{1}{a^2 + \delta_1^2} \times \nn \\
    &\times&
    \left\{
        \frac{a}{2}
        \ln \br{\frac{\br{a+b}^2 + \delta_2^2}{\br{a-b}^2 + \delta_2^2}}
        \mp
        \delta_1
        \arctan\br{\frac{2 b \delta_2}{a^2-b^2+\delta_2^2}}
    \right\}.
\end{eqnarray}

The limit of theses integrals for small values of $|\vec {\cal Q}|$ is:
$$\left .I_\pm \right |_{|\vec {\cal Q}|\to 0}= \frac{4}{\pi} \int\limits_{-M\sqrt{\tau}}^{M\sqrt{\tau}} d \kappa_0
\int\limits_0^{M\sqrt{\tau-k_0^2/M^2}}
\displaystyle\frac{ d k~k^2(a^2\mp \delta_1\delta_2)}
{(a^2+\delta_2^2)(a^2+\delta_1^2)}.$$

One can see that in  the limiting case: $\epsilon\to 0$ and $\theta\to \pi$, $I_\pm$ is finite, whereas for $\epsilon\to 1$ and $\theta\to 0$, $I_\pm \to 0$. This is responsible for the behavior of the two photon exchange contribution as a function of $\epsilon$.



\begin{thebibliography}{99}

\bibitem{MaxT}
  L.~C.~Maximon and J.~A.~Tjon,
  Phys.\ Rev.\ C {\bf 62}, 054320 (2000).
\bibitem{KMF}
  E.~A.~Kuraev, N.~P.~Merenkov and V.~S.~Fadin,
  Sov.\ J.\ Nucl.\ Phys.\  {\bf 47}, 1009 (1988)
  [Yad.\ Fiz.\  {\bf 47}, 1593 (1988)].


\bibitem{AAM}
  A.~Afanasev, I.~Akushevich and N.~Merenkov,
  Phys.\ Rev.\ D {\bf 64}, 113009 (2001).

\bibitem{DKSV}
S. Dubni\c cka, E. Kuraev, M. Se\c canski and A. Vinnikov,
hep-ph/0507242.

\bibitem{Mo69} L. W. Mo and Y. S. Tsai, Rev. Mod. Phys. {\bf 41}, 205 (1969).


\bibitem{KF}
  E.~A.~Kuraev and V.~S.~Fadin,
  Sov.\ J.\ Nucl.\ Phys.\  {\bf 41}, 466 (1985)
  [Yad.\ Fiz.\  {\bf 41}, 733 (1985)];
  M.~Skrzypek,
  Acta Phys.\ Polon.\ B {\bf 23}, 135 (1992).

\bibitem{Gu73}  J. Gunion and L. Stodolsky,  Phys. Rev. Lett. {\bf 30}, 345
(1973); V. Franco,  Phys. Rev. D {\bf 8}, 826 (1973);  V. N. Boitsov, L.A. Kondratyuk and V.B. Kopeliovich, Sov. J.
Nucl. Phys {\bf 16}, 287 (1973);  F. M. Lev, Sov. J. Nucl. Phys. {\bf 21}, 45 (1973);
\bibitem{Bl71}
R. Blankenbecker and J. Gunion, Phys.  Rev. D {\bf 4}, 718
(1971).
\bibitem{Fr99}
  V.~V.~Frolov {\it et al.},
  Phys.\ Rev.\ Lett.\  {\bf 82}, 45 (1999).


\bibitem{Re99} M. P. Rekalo, E. Tomasi-Gustafsson and D. Prout, Phys. Rev.
{\bf  C60}, 042202 (1999).

\bibitem{Re68} A. Akhiezer and M. P. Rekalo, Dokl. Akad. Nauk USSR, {\bf 180},
1081 (1968); Sov. J. Part. Nucl. {\bf 4}, 277 (1974).
\bibitem{Jo00}
M. K. Jones {\it et al.}, Phys. Rev. Lett. 84 (2000) 1398;
O. Gayou {\it et al.}, Phys. Rev. Lett. 88 (2002) 092301;
  V.~Punjabi {\it et al.},
  Phys.\ Rev.\ C 71 (2005) 055202.
\bibitem{Ar04}
  I.~A.~Qattan {\it et al.},
  Phys.\ Rev.\ Lett.\  {\bf 94}, 142301 (2005).
\bibitem{ET04}
  E.~Tomasi-Gustafsson and G.~I.~Gakh,
  Phys.\ Rev.\ C {\bf 72}, 015209 (2005).

\bibitem{Bl05}
  P.~G.~Blunden, W.~Melnitchouk and J.~A.~Tjon,
  Phys.\ Rev.\ C {\bf 72}, 034612 (2005).
  A.~V.~Afanasev, S.~J.~Brodsky, C.~E.~Carlson, Y.~C.~Chen and M.~Vanderhaeghen,
  Phys.\ Rev.\ D {\bf 72}, 013008 (2005).
\bibitem{bvv}
  E.~A.~Kuraev, V.~V.~Bytev, Y.~M.~Bystritskiy and E.~Tomasi-Gustafsson,
  %
   Phys.\ Rev.\ D\  {\bf 74}, 013003 (2006).
\bibitem{Ko05}
  S.~Kondratyuk, P.~G.~Blunden, W.~Melnitchouk and J.~A.~Tjon,
  Phys.\ Rev.\ Lett.\  {\bf 95}, 172503 (2005).
\bibitem{AR77}
A.I. Akhiezer and M.P. Rekalo "Electrodynamics of Hadrons", Kiev,
Naukova Dumka, 1977.
\bibitem{Ja66} T. Janssens, R. Hofstadter, E. B. Hughes and M. R. Yerian,
Phys. Rev. {\bf 142}, 922 (1966).
\bibitem{An94}
L.~Andivahis {\it et al.},
Phys.\ Rev.\ D {\bf 50}, 5491 (1994).
\bibitem{Ch04}
M.~E.~Christy {\it et al.}  [E94110 Collaboration],
Phys.\ Rev.\ C {\bf 70}, 015206 (2004).

\bibitem{Af00}
  A.~V.~Afanasev, I.~Akushevich and N.~P.~Merenkov,
  Phys.\ Rev.\ D {\bf 65}, 013006 (2002).
\bibitem{ETG04}
  E.~Tomasi-Gustafsson,
  arXiv:hep-ph/0412216.

\bibitem{Ku06}
  E.~A.~Kuraev, M.~Secansky and E.~Tomasi-Gustafsson,
  Phys.\ Rev.\ D {\bf 73}, 125016 (2006).

\bibitem{Bytev} V. Bytev {\it et al.}, in preparation.

\end{thebibliography}
\end{document}